\documentclass[sigconf]{acmart}
\AtBeginDocument{%
  }

\setcopyright{acmlicensed}
\copyrightyear{2026}
\acmYear{2026}
\acmDOI{XXXXXXX.XXXXXXX}

\acmISBN{978-1-4503-XXXX-X/2018/06}

\usepackage{amsmath} 
\usepackage{amsfonts} 
\usepackage{mathrsfs}
\usepackage{enumitem}
\usepackage{booktabs}  
\usepackage{tcolorbox}
\usepackage{multirow} 
\usepackage{graphicx}  
\usepackage{amsmath}  
\usepackage{mathrsfs}  
\usepackage{subcaption}
\usepackage{caption}
\usepackage{subcaption}

\usepackage{algorithm}
\usepackage{algorithmic}
\allowdisplaybreaks

\begin{document}

\title{Masked Diffusion Generative Recommendation}

\author{Lingyu Mu}
\authornotemark[1]
\affiliation{
  \institution{Alibaba International Digital Commerce Group}
  \city{Beijing} 
  \state{} 
  \country{China}
}
\email{mulingyu@iie.ac.cn}

\author{Hao Deng}
\orcid{0009-0002-6335-7405}
\authornote{Contributed equally to this research.} 
\affiliation{%
  \institution{Alibaba International Digital Commerce Group}
   \city{Beijing} 
   \state{} 
   \country{China}
}
\email{denghao.deng@alibaba-inc.com}

\author{Haibo Xing}
\orcid{0009-0006-5786-7627}
\affiliation{%
  \institution{Alibaba International Digital Commerce Group}
  \city{Hangzhou} 
  \state{} 
  \country{China}
}
\email{xinghaibo.xhb@alibaba-inc.com}

\author{Jinxin Hu}
\authornote{Corresponding authors.}
\orcid{0000-0002-7252-5207}
\affiliation{
  \institution{Alibaba International Digital Commerce Group}
  \city{Beijing} 
  \state{} 
  \country{China}
}
\email{jinxin.hjx@alibaba-inc.com}

\author{Yu Zhang}
\orcid{0000-0002-6057-7886}
\affiliation{
  \institution{Alibaba International Digital Commerce Group}
  \city{Beijing} 
  \state{} 
  \country{China}
}
\email{daoji@alibaba-inc.com}

\author{Xiaoyi Zeng}
\orcid{0000-0002-3742-4910}
\affiliation{
  \institution{Alibaba International Digital Commerce Group}
  \city{Hangzhou} 
  \state{} 
  \country{China}
}
\email{yuanhan@taobao.com}

\author{Jing Zhang}
\authornotemark[2]
\orcid{0000-0002-7252-5207}
\affiliation{
  \institution{Wuhan University}
  \city{Wuhan} 
  \state{} 
  \country{China}
}
\email{jingzhang.cv@gmail.com}

\begin{abstract}
Generative recommendation (GR) typically first quantizes continuous item embeddings into multi-level semantic IDs (SIDs), and then generates the next item via autoregressive decoding. Although existing methods are already competitive in terms of recommendation performance, directly inheriting the autoregressive decoding paradigm from language models still suffers from three key limitations: (1) autoregressive decoding struggles to jointly capture global dependencies among the multi-dimensional features associated with different positions of SID; (2) using a unified, fixed decoding path for the same item implicitly assumes that all users attend to item attributes in the same order; (3) autoregressive decoding is inefficient at inference time and struggles to meet real-time requirements. To tackle these challenges, we propose MDGR, a \textbf{M}asked \textbf{D}iffusion \textbf{G}enerative \textbf{R}ecommendation framework that reshapes the GR pipeline from three perspectives: codebook, training, and inference. (1) We adopt a parallel codebook to provide a structural foundation for diffusion-based GR. (2) During training, we adaptively construct masking supervision signals along both the temporal and sample dimensions. (3) During inference, we develop a warm-up–based two-stage parallel decoding strategy for efficient generation of SIDs. Extensive experiments on multiple public and industrial-scale datasets show that MDGR outperforms ten state-of-the-art baselines by up to 10.78\%. 
Furthermore, by deploying MDGR on a large-scale online advertising platform, we achieve a 1.20\% increase in revenue, demonstrating its practical value.
\end{abstract}

\vspace{-3pt}
\ccsdesc[500]{Information systems~Retrieval models and ranking}
\vspace{-3pt}
\keywords{Generative Recommendation, Masked Diffusion Model}


\maketitle


\begin{figure}[t]
  \centering
  \includegraphics[width=1.0    \linewidth]{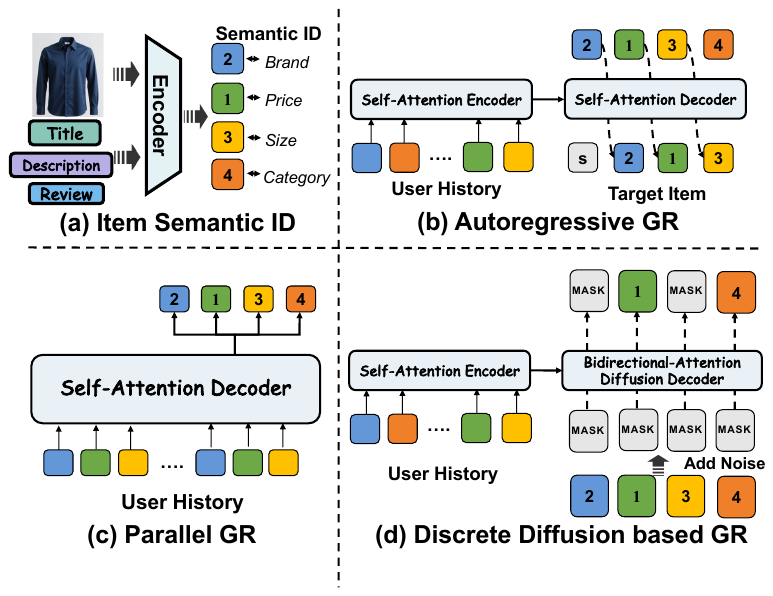}
  \caption{(a) The codebook quantizes the multimodal information of an item into a sequence of semantic tokens, i.e., SIDs. (b) Autoregressive GR generates SIDs in a fixed left‑to‑right order. (c) Parallel GR generates all tokens in a single step. (d) Our masked diffusion GR denoises multiple positions in parallel, flexibly filling tokens without a fixed order.}
  \label{fig1}
  \vspace{-0.5cm} 
\end{figure}

\section{Introduction}
In recent years, generative recommendation (GR) based on semantic IDs (SIDs) has attracted extensive attention in both academia and industry \cite{tiger, cobra}. Unlike traditional recommender systems (RSs) \cite{wang2021survey, wang2024rethinking, lin2024enhancing,wang2025home,mu2025trust,deng2025csmf}, which assign each item a unique ID and learn a dedicated embedding for it, GR typically leverages pre-trained models to map an item’s content features (such as title, description, and image) into a continuous semantic space, and then applies vector quantization \cite{gray1984vector} to compress each item into a set of discrete tokens, i.e., SIDs. As illustrated in Figure \ref{fig1}(a), different tokens in a SID often correspond to the item’s different characteristics \cite{rpg, zhou2025onerec, mu2025synergistic}, such as category, brand, and price, while the combination of all tokens determines the item’s overall position in the semantic space. 
By indexing items with shared semantic tokens instead of IDs, GR represents a large item corpus with a compact token vocabulary \cite{wang2024learnable}, thereby improving scalability and memory efficiency.

Existing GR can be roughly divided into two types: autoregressive decoding with residual codebooks and single-step decoding with parallel codebooks. The former follows the modeling paradigm of language models. It constructs hierarchical SIDs for items via multi-level residual quantization \cite{lee2022autoregressive}, and then generates tokens one by one from left to right, achieving competitive performance. However, this paradigm has two limitations: 
(1) The objective of GR is whether the final generated SIDs corresponds to the target item \cite{wang2023generative, zhang2024generative}, not the specific generation order. Yet autoregressive models can only condition on the left prefix when predicting each token \cite{vaswani2017attention}, limiting their ability to enforce global consistency across tokens \cite{rpg}. (2) Users’ click interests are inherently heterogeneous \cite{xing2025reg4rec, yao2016things, zhang2016modeling}: different users may attend to an item’s attributes in different orders. A fixed, user-agnostic decoding path implicitly assumes a shared attention order, contradicting this heterogeneity. 
To better model global consistency, single-step decoding with parallel codebooks decomposes item semantics into multiple sub-codebooks and predicts all codewords simultaneously in one forward pass, enabling synchronized modeling of semantic dimensions and improving overall consistency compared with autoregressive decoding.
Nevertheless, this paradigm still has two drawbacks. (1) The one-shot decision process tends to overlook finer-grained correlations and constraints among local attributes. (2) Its decoding process still fails to resolve the mismatch between heterogeneous user interests and the fixed decoding order. Therefore, under the premise of using parallel codebooks, we aim to design a new decoding paradigm that meets the following three requirements: (1) \textbf{Order-agnostic:} the generation order of semantic IDs can be adjusted according to the user’s interest structure. (2) \textbf{Multi-step refinement:} SIDs can be progressively refined over multiple iterations to improve generation accuracy. (3) \textbf{Parallel generation:} multiple positions are updated simultaneously at each step to ensure efficiency.

To address the above issues, inspired by recent advances in discrete diffusion models \cite{lou2023reflected,graves2023bayesian,lin2023text,xue2024unifying,zhang2025target}, we explore incorporating the idea of masked diffusion \cite{austin2021structured} into GR. As shown in Figure \ref{fig1}(d), the masked diffusion model treats generation as a multi-step masking–denoising process. Starting from a sequence of \texttt{[MASK]} tokens, it iteratively selects positions to recover into concrete SIDs using bidirectional attention. This order-free mechanism supports gradual refinement of token semantics and parallel denoising of multiple positions at each step, reducing decoding steps while maintaining a flexible decoding path.
However, directly applying discrete diffusion to recommendation faces several challenges. 
(1) Standard masked diffusion models use a fixed noise schedule and inject noise uniformly at random \cite{peebles2023scalable}. In GR, however, SID tokens differ in importance and difficulty, so such SID-agnostic masking weakens modeling of the true interest structure.
(2) Existing diffusion methods generate only a single sequence \cite{bao2023all}, whereas GR retrieval requires a mechanism to produce diverse candidates like beam search. Since beam search is designed for autoregressive step‑wise decoding and cannot be directly used for parallel decoding, we need a new decoding mechanism for diffusion‑based GR.

To tackle the above challenges, we propose MDGR, a \textbf{\underline{M}}asked \textbf{\underline{D}}iffusion \textbf{\underline{G}}enerative \textbf{\underline{R}}ecommendation framework that reshapes both the training and inference procedures. Specifically, \textbf{in the  training stage}, we dynamically control the noise distribution along both the temporal and sample dimensions to provide more effective supervision for multi-step generation:
\begin{itemize}[noitemsep, topsep=0pt, leftmargin=*]
    \item \textbf{Temporal Dimension.} Inspired by curriculum learning \cite{bengio2009curriculum}, we further propose a \textbf{global curriculum noise scheduling} strategy, where the masking ratio is gradually increased as training progresses, exposing the model to increasingly challenging reconstruction tasks.
    \item \textbf{Sample Dimension.} Given the sampled number of masks, we further propose a \textbf{history-aware masking allocation} strategy that prioritizes masking semantic tokens that are rare in the user’s history, thereby exposing the model to harder examples.
\end{itemize}
\textbf{In the inference stage}, we adopt a warm-up–based two-stage parallel decoding: a few steps of single-position decoding are first used to stabilize key semantic anchors, after which we switch to parallel prediction over multiple token groups, which can be combined with beam search \cite{vaswani2017attention} for efficient candidate SID generation.

We conduct extensive experiments on two public datasets and one industrial dataset to evaluate MDGR, and compare it with ten state-of-the-art (SOTA) baselines (\textit{e.g.}, TIGER \cite{tiger}, Cobra \cite{cobra}). MDGR achieves the best performance across all comparison settings, with up to 10.78\% improvement in performance. Moreover, we deploy MDGR on a large commercial advertising platform. 
Online A/B testing shows that advertising revenue increases by 1.20\% and gross merchandise volume (GMV) increases by 3.69\%. 

Our contributions can be summarized as follows:
\begin{itemize}[noitemsep, topsep=0pt, leftmargin=*]
    \item We propose MDGR, a masked diffusion GR that models SID generation as a masking–denoising process, enabling bidirectional modeling across token dimensions and parallel prediction. 
    \item During training, we design a dynamic global noise scheduling strategy along both the temporal and sample dimensions, progressively constructing hard missing-position supervision signals tailored to users’ heterogeneous click interests. During inference, we develop a warm-up–based two-stage parallel decoding strategy to efficiently generate SIDs.
    \item We achieve SOTA results on multiple datasets, with 7.17\%–10.78\% improvement over both discriminative and generative baselines, demonstrating that diffusion GR is an effective new paradigm.

\end{itemize}

\section{Related Work}

\subsection{Generative Recommendation}
GR maps item content (\textit{e.g.}, text, images) into a continuous semantic space and then applies vector quantization to obtain discrete SIDs \cite{hua2023index,hou2024bridging}. TIGER \cite{tiger} is the earliest SID‑based GR model, using RQ‑VAE to quantize text features and an autoregressive Transformer to generate SIDs token by token. Cobra \cite{cobra} augments SIDs with continuous vectors and alternates between generating discrete codes and continuous embeddings to bridge generation and retrieval. RPG \cite{rpg} further improves codebook expressiveness via optimized product quantization \cite{ge2013optimized}. It splits an item’s semantic vector into multiple subspaces and quantizes each independently to form a longer SID with finer‑grained semantics. Despite these advances in SID and codebook design, most GR methods still follow a standard language‑modeling paradigm. They use next‑token prediction as supervision and generate SIDs autoregressively in a fixed order. This limits their ability to model multidimensional user interest heterogeneity and to meet the decoding‑efficiency requirements of recommendation scenarios.

\subsection{Discrete Diffusion Models}
Diffusion models initially achieved great success in continuous spaces (such as image generation) by gradually injecting Gaussian noise into data in the forward process and learning a reverse denoising process to approximate the data distribution \cite{lou2023discrete,shi2024simplified,lin2025order,yang2023diffusion}. To extend this idea to discrete sequences, a series of discrete diffusion models (DDMs) has emerged in recent years. These methods typically define noise as a form of discrete perturbation. For example, masked diffusion models (MDMs) \cite{sahoo2024simple,ou2024your} construct a Markov diffusion process in a finite state by replacing tokens with a special \texttt{[MASK]} symbol or randomly substituting them from the vocabulary, and then training a model to restore masked positions at different noise levels. LLaDA \cite{nie2025large} is the first diffusion-based language model. By masking data and restoring it in the reverse process, LLaDA  achieves performance comparable to autoregressive models and alleviates the reversal curse problem. LLaDA 1.5 \cite{zhu2025llada} further introduces reinforcement learning, significantly improving the performance of MDM alignment.
Most existing DDMs target image or text generation \cite{chang2022maskgit,chang2023muse,you2025effective}, whereas SIDs in GR are unordered multi-token item identifiers with distinct structural constraints. Inspired by multi-step noising and gradual denoising, we adapt discrete diffusion to SIDs by designing training-time corruption schemes, enabling the model to complete full SID sequences under varying information completeness. 
Recently, concurrent studies also apply diffusion to GR \cite{baidu, ruc}, mainly by directly adopting generic MDMs architectures with minimal adaptation to recommendation. In contrast, we introduce recommendation-specific adaptations to the diffusion process and empirically validate their effectiveness in Sec.~\ref{sec:Ablation}.

\section{Preliminaries}
\subsection{Generative Recommendation}
Let $\mathcal{I}$ be the item set and $\mathcal{U}$ be the user set. For each user $u \in \mathcal{U}$, the historical interaction sequence is $s_u=(i_1, i_2,...,i_T)$. GR treats recommendation as a generation problem: given \(s_u\), the model generates a discrete identifier representing the target item. Specifically, we define $L$ discrete vocabularies $\{V_\ell\}_{\ell=1}^L$. 
Each $V_\ell$ is associated with a codebook matrix 
$C_\ell \in \mathbb{R}^{|V_\ell| \times d_\ell}$ that stores the 
codeword of each index. Given these codebooks, each item $i$ is represented by a length‑$L$ token sequence $c_i = (c_i^1, \ldots, c_i^L),$ where each token $c_i^\ell \in V_\ell$ is the index of a codeword in the $\ell$‑th codebook $C_\ell$. The objective of GR is to learn the conditional distribution $p_{\theta}(c|s_u)$, i.e., generating a SID $c$ conditioned on the user history $s_u$, and then mapping it back to a concrete item $i$ via codebook-based retrieval. During training, given $s_u$ and the next target item $i^+$, we use $c_{i^+}$ 
as supervision and maximize 
$\max_{\theta}\mathbb{E}_{(u,i^+)}[\log p_{\theta}(c_{i^+}\mid s_u)]$. Under the autoregressive paradigm, this conditional distribution is:
\begin{small}
 \begin{equation}
p_{\theta}(c_{i^+}\mid s_u)= \prod_{\ell=1}^{L} p_{\theta}\big(c_{i^+}^{\ell} \mid c_{i^+}^{<\ell}, s_u\big),
\label{eq:factorize}
\end{equation}   
\end{small} and the model is trained via next-token prediction, i.e., by minimizing the negative log-likelihood of the ground-truth target SIDs:
\begin{small}
\begin{equation}
\mathcal{L}_{\mathrm{AR}}= -\sum_{\ell=1}^{L} \log p_{\theta}\big(c_{i^+}^{\ell} \mid c_{i^+}^{<\ell}, s_u\big).
\end{equation}   
\end{small}
\subsection{Discrete Diffusion Models}
DDMs define a Markov chain \cite{sahoo2024simple} over a discrete space that gradually adds noise to a target sequence and then learns the reverse process to recover it. Let $z \in V^L$ be a discrete sequence of length $L$. The forward process defines a Markov chain from $t=0$ to $t=T$:
\begin{equation}
q(\mathbf{z}^{0:T} \mid \mathbf{z}^{0})= \prod_{t=1}^{T} q(\mathbf{z}^{(t)} \mid \mathbf{z}^{(t-1)}),
\end{equation}
where $z^{(0)}=z$, and $z^{(T)}$ approaches a high‑noise discrete‑state distribution. Correspondingly, the reverse process reconstructs $z^{(0)}=z$ from any intermediate noisy state $z^{(t)}$:
\begin{equation}
p_{\theta}(\mathbf{z}^{0:T}) = \pi(\mathbf{z}^{(T)}) \prod_{t=1}^{T} p_{\theta}\big(\mathbf{z}^{(t-1)} \mid \mathbf{z}^{(t)}, \mathbf{y}\big),
\end{equation}
where $\mathbf{y}$ is the condition, $\pi(\mathbf{z}^{(T)})$ is usually set to a simple prior distribution (\textit{e.g.}, the fully masked state), and $p_{\theta}$ is the parameterized model to be learned.
In our setting, $z$ corresponds to the SID of the target item and $\mathbf{y}$ denotes the user interaction sequence. We instantiate $q(\cdot)$ as a masking-based corruption process over SIDs and train $p_\theta$ to predict the ground-truth tokens on the corrupted positions. Concretely, let $x_0$ be the target item’s SID, $x_\tau \sim q_\tau(\cdot \mid x_0)$ be the noisy SID at timestep $\tau$, and $\mathcal{M}_\tau$ be the set of masked positions. The model takes $(x_\tau, s_u, \tau)$ as input and outputs a distribution over the codebook at each position. The training objective is the cross-entropy loss on masked positions:
\begin{small}
\begin{equation}
\mathcal{L}_{\mathrm{DDM}}= \mathbb{E}_{(u,i^+)} \, \mathbb{E}_{\tau \sim p(\tau)} \, \mathbb{E}_{x_{\tau} \sim q_{\tau}(\cdot \mid x_0)}\Bigg[ - \sum_{\ell \in \mathcal{M}_{\tau}} \log p_{\theta}\big(x_0^{\ell} \mid x_{\tau}, \tau, s_u\big) \Bigg],  
\end{equation}    
\end{small}
where $(u, i^+)$ denotes a positive user-item pair, $x^l_0$is the ground-truth token of the target SID at position $\ell$, $p(\tau)$ is the sampling distribution over timesteps, and $q_{\tau}(\cdot \mid x_0)$ is the forward noising distribution at timestep $\tau$ conditioned on $x_0$.

\begin{figure*}[ht]
  \centering
  \includegraphics[width=1.0\textwidth]{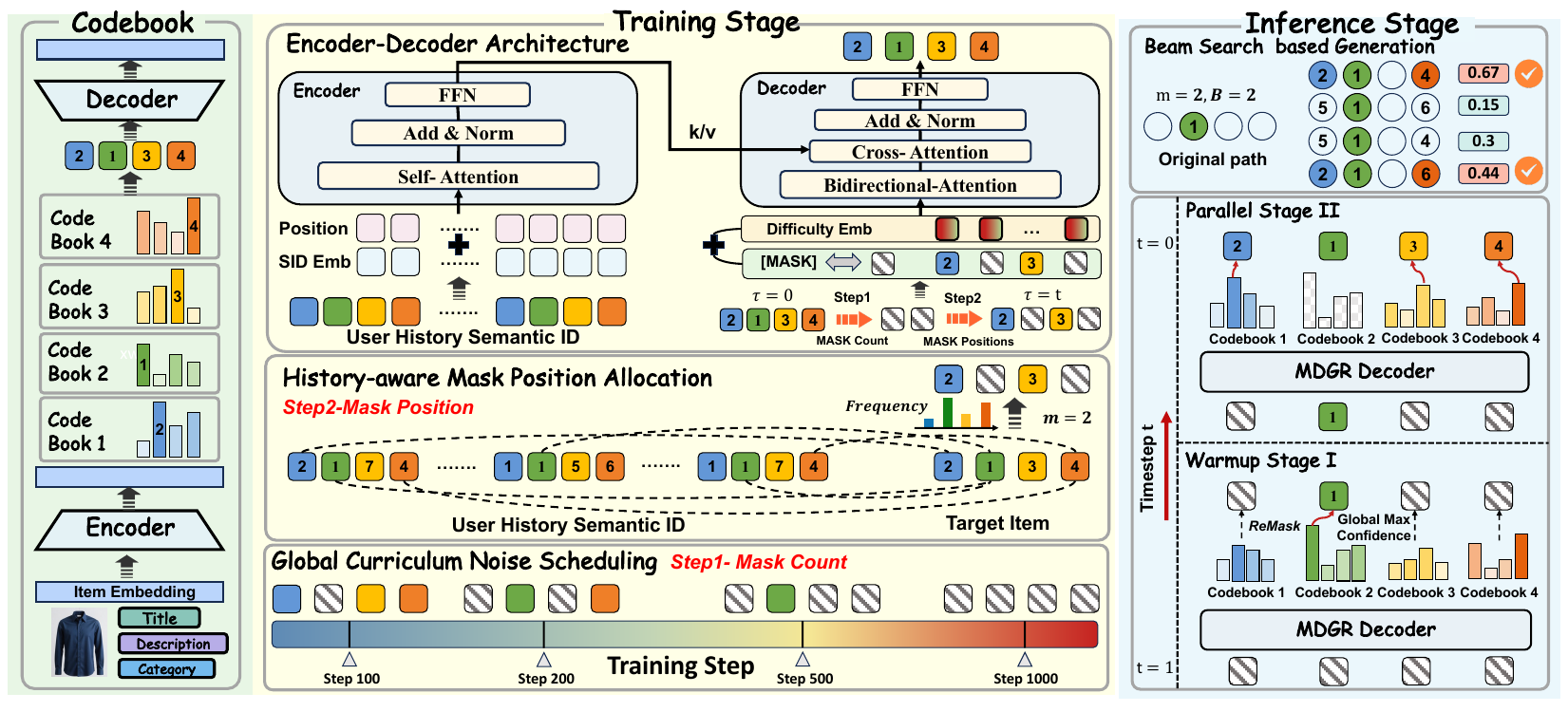 } 
  \caption{The overview of MDGR: a masked diffusion generative recommendation. (1) Codebook: we adopt an OPQ-based parallel codebook to obtain SIDs for items. (2) Training: we use an encoder–decoder architecture. Based on the current training stage, we first determine the number of masks via global curriculum noise scheduling, and then derive the masked positions for each sample using history-aware mask allocation. (3) Inference: we employ a warm-up–based two-stage parallel decoding strategy, combined with beam search, to jointly generate Top‑B candidate items across multiple codebooks.}
  \label{fig:2}
\end{figure*}

\section{Method}
\subsection{Overview}
This section introduces MDGR. As shown in Figure~\ref{fig:2}, it consists of three parts. 
\textbf{(i) Parallel codebook.} Each item's semantic vector is split into subspaces and quantized independently, yielding a multi-token SID for parallel semantic modeling. 
\textbf{(ii) Training.} We view SID generation as a masked diffusion process, where a temporal curriculum gradually increases the masking ratio, and token-preference--based masking with a difficulty vector enables global difficulty control and sample-specific noise. 
\textbf{(iii) Inference.} From a fully masked SID, we adopt a two-stage decoding strategy. The warm-up stage first captures coarse semantics, and the subsequent parallel stage accelerates multi-position prediction. Finally, beam search yields complete SIDs, which are mapped back to items for Top-$K$ recommendation.

\subsection{Parallel Codebook}

Existing works commonly adopt residual quantization \cite{tiger,lee2022autoregressive}, where codebooks are constructed by sequentially quantizing residual vectors. Although this design achieves a high compression rate, lower‑level tokens only encode residual semantics conditioned on higher‑level tokens, which naturally biases the generation process toward a fixed hierarchical order. This conflicts with our goal of leveraging bidirectional context and parallel denoising. Therefore, we instead build item SIDs using OPQ-based parallel codebooks, where each token resides in a relatively independent semantic subspace and all tokens jointly determine the item.

Concretely, given an item content representation $\mathbf{e}_i \in \mathbb{R}^d$ from a pretrained encoder, we project it into $L$ subspaces:
\begin{equation}
\mathbf{e}_i^{\ell} = f_\ell(\mathbf{e}_i) \in \mathbb{R}^{d_\ell}, \quad \ell = 1,\ldots,L,
\end{equation}
where $f_\ell$ is a linear projection layer for the $\ell$‑th subspace. For each $\ell$, we maintain a codebook $C_\ell \in \mathbb{R}^{|V_\ell|\times d_\ell}$. We then assign $\mathbf{e}_i^{\ell}$ to its nearest codeword by:
\begin{equation}
c_i^{\ell} = \arg\min_{j \in V_\ell} \left\lVert \mathbf{e}_i^{\ell} - C_\ell[j] \right\rVert_2^{2},
\end{equation}
where $C_\ell[j]$ denotes the $j$-th codeword and $c_i^{\ell}$ is its index in $V_\ell$. Collecting the indices from all subspaces yields the parallel SID
$c_i = (c_i^1,\dots,c_i^L)$ for item $i$, which naturally supports bidirectional modeling and independent masking across semantic dimensions.

\subsection{Training Stage}
We adopt a standard encoder–decoder architecture, where the encoder encodes the user’s interaction history as external context for the decoder. On the decoder side, we replace the traditional causal self-attention with bidirectional attention to accommodate the discrete diffusion process. Beyond the architecture, we dynamically control the noise distribution along both the temporal and sample dimensions to adapt to different training stages and sample characteristics. Overall, given a user history $s_u$ and the target SID $c_{i^+}$ of the target item, we first sample the number of masked positions (temporal dimension), then select which tokens to mask (sample dimension). The model is trained to reconstruct the original tokens from the partially masked $\tilde{c}_{i^+}$, with a difficulty-aware vector indicating the current noise level.

\subsubsection{Temporal dimension: global curriculum noise scheduling strategy}
To enable the model to gradually learn reconstruction under noise, we design a global curriculum noise schedule based on training progress. At the beginning of training, the model capacity is limited, so we assign easier instances with fewer masked positions and gradually increase the masking rate as training proceeds. Specifically, let $N$ be the total number of training steps and $n$ the current step. We first normalize the training progress as $\tau = \min(1, n/N)$, and map $\tau$ to a noise difficulty scalar via a smooth cosine schedule inspired by diffusion models. We then apply a power transformation to this scalar to obtain the stretched difficulty $\delta$.

\begin{small}
\begin{equation}
\delta
= \biggl(\sqrt{1 - \cos^{2}\!\Big(\frac{\pi}{2}(1-\tau)\Big)}\biggl)^{\gamma},
\label{eq:stretched_difficulty}
\end{equation}
\end{small}
where $\gamma$ is a hyperparameter that controls how fast $\delta$ decreases from 1 to 0. 

Given the maximum number of masked positions $L$, we aim to decide at how many positions to mask based on the current difficulty level $\delta \in [0,1]$. 
Early in training we prefer easier cases with fewer masks, while later in training we gradually shift towards harder cases with more masks. Formally, we consider all possible mask counts $k \in \{1,2,\dots,L\}$ and construct a $\delta$-dependent distribution over them.
We first define two monotonic scoring functions over $k$: $f_{\text{low}}(k)$, which decreases with $k$ and thus favors small mask counts, and $f_{\text{high}}(k)$ which increases with $k$ and thus favors large mask counts.\footnote{In practice, we use simple linear functions, \textit{e.g.}, $f_{\text{low}}(k) = L+1-k$ and $f_{\text{high}}(k) = k$, but other monotonic forms also work.} We then interpolate between these two scores using $\delta$:
\begin{small}
\begin{equation}
s(k) = (1 - \delta)\,f_{\text{high}}(k) + \delta\,f_{\text{low}}(k), \quad k = 1,\dots,L.
\end{equation}
\end{small}
When $\delta$ is large (early training), $s(k)$ is dominated by $f_{\text{low}}(k)$ and assigns higher scores to small $k$. As $\delta$ decreases (later training), the influence of $f_{\text{high}}(k) = k$ grows, and the scores shift towards larger $k$. We normalize these scores to obtain a probability distribution over the mask count:
\begin{small}
    \begin{equation}
        \mathbf{P}_{time}(k) = \frac{s(k)}{\sum_{j=1}^{L} s(j)}, \quad k = 1,\dots,L.
    \end{equation}
\end{small}
Finally, we sample the number of masked positions for each training instance as $k \sim P_{time}(k)$. Thus, as $\delta$ decreases from $1$ to $0$ over training, the probability mass of $P_{time}(k)$ gradually moves from small to large $k$, implementing a curriculum from easy to hard masking patterns.

To make the decoder aware of how much noise is applied at each step, we introduce a learnable difficulty-aware embedding. 
Specifically, given the sampled number of masked positions $k$, we treat $k$ as a discrete difficulty index and retrieve a difficulty-aware embedding $\mathbf{d}_{k}$ from an embedding table $\mathbf{D} \in \mathbb{R}^{L \times D}$. We then add $\mathbf{d}_{k}$ to every input token embedding, providing a global conditioning signal on instance difficulty and stabilizing curriculum training.
 
\subsubsection{Sample dimension: history-aware mask position allocation}
After determining the number of masked positions for each sample, we further decide where to place these masks so that training better aligns with the user’s personalized interest structure. To this end, we propose a history‑aware mask allocation strategy that dynamically selects positions to mask according to how frequently each token of the target item appears in the user’s history.
Specifically, for user $u$ and target item $i^+$, let $c_{i^+}=(c_{i^+}^1,c_{i^+}^2,...,c_{i^+}^L)$ be the target SID and $s_u$ be the user’s interaction history. We count how many times the tokens at each position of the $c_{i^+}$ appear in the history:
\begin{equation}
 f^{\ell} = \#\{\, j \in s_u \mid c_j^{\ell} = c_{i^+}^{\ell} \},
\qquad \ell = 1,\ldots,L, 
\end{equation}
where $\#\{\cdot\}$ denotes the number of elements in the set. Based on this, we construct a normalized distribution that reflects prediction difficulty, where tokens with lower frequency are regarded as harder to predict. Let $\mathbf{P}_{pos} = (p^{1}_{pos}, \dots, p^{L}_{pos})$ denote the resulting position-wise sampling distribution, defined as:
\begin{small}
\begin{equation}
w^{\ell} = \frac{1}{f^{\ell} + \varepsilon}, \qquad
\mathbf{P}^{\ell}_{pos} = \frac{w^{\ell}}{\sum_{m=1}^{L} w^{m}}, \qquad
\ell = 1,\ldots,L,
\end{equation}
\end{small}
where $\varepsilon$ is a small positive constant. 
Given the mask count $k$ for a sample, we sample a mask position set $\mathcal{M} \subseteq \{1,\dots,L\}$ with $|\mathcal{M}| = k$ according to the position-wise distribution $\mathbf{P}_{\text{pos}}$, i.e., $\mathcal{M} \sim \mathbf{P}_{\text{pos}}$.
With this strategy, more mask budget is assigned to semantic dimensions that appear less frequently in the user’s history and are thus harder to predict, focusing supervision on challenging semantics.
We then replace tokens at positions in $\mathcal{M}$ with a shared placeholder symbol \texttt{[MASK]}, while keeping tokens at unmasked positions as their original tokens:
\begin{small}
\begin{equation}
\tilde{c}^{\ell}_{i+} =
\begin{cases}
\text{[MASK]}, & \ell \in \mathcal{M},\\[2pt]
c^{\ell}_{i+},    & \ell \notin \mathcal{M},
\end{cases}
\qquad \ell = 1,\ldots,L.    
\end{equation}
\end{small}

Overall, MDGR can be viewed as a discrete-time absorbing Markov process over the target SID $c_{i^+} = (c_{i^+}^1,\dots,c_{i^+}^L)$. Let $\{c_t\}_{t=0}^{T}$ denote the latent corrupted SIDs with $c_0 = c_{i^+}$ and $c_t^\ell \in V_\ell\ \cup\{\text{[MASK]}\}$.
At each Markov timestep $t$, every semantic dimension $\ell$ independently
either stays unchanged or is absorbed into the \texttt{[MASK]} state according to the following single-dimension transition kernel:
\begin{small}
\begin{equation}
q(c_{t+1}^\ell\mid c_t^\ell, u, i^+)
=\begin{cases}
1, &c_t^\ell=\text{[MASK]},\ c_{t+1}^\ell=\text{[MASK]},\\
\alpha_{t,\ell}(u,i^+), &c_t^\ell \neq \text{[MASK]},\ c_{t+1}^\ell = \text{[MASK]},\\
1 - \alpha_{t,\ell}(u,i^+), &c_t^\ell \neq \text{[MASK]},\ c_{t+1}^\ell = c_t^\ell,\\
0, &\text{otherwise},
\end{cases}
\end{equation}
\end{small}
where $\alpha_{t,\ell}(u,i^+) \in [0,1]$ is the masking rate of semantic dimension $\ell$ at timestep $t$. The full transition kernel factorizes over dimensions as
\(q(c_{t+1} \mid c_t,u,i^+) = \prod_{\ell=1}^L q(c_{t+1}^\ell \mid c_t^\ell,u,i^+)\). 
In practice, these effective masking rates \(\alpha_{t,\ell}(u,i^+)\) are not parameterized explicitly; instead, they are implicitly induced by our global curriculum noise schedule $\mathbf{P}_{\text{time}}$ and
history-aware mask allocation $\mathbf{P}_{pos}$.

At each training step we sample a mask count $k \sim \mathbf{P}_{\text{time}}(k)$ and a mask position set $\mathcal{M} \sim \mathbf{P}_{\text{pos}}(\cdot \mid k,u,i^+)$ to construct the noised SID with positions in $\mathcal{M}$. Given user history $s_u$ and difficulty-aware embedding $\mathbf{d}_{k}$, the decoder acts as a denoiser that reconstructs masked tokens, yielding a masked training objective over SIDs:
\begin{small}
\begin{equation}
\mathcal{L}
= \mathbb{E}_{(u,i^+)}
  \mathbb{E}_{k \sim \mathbf{P}_{\text{time}}(\cdot)}
  \mathbb{E}_{\mathcal{M} \sim \mathbf{P}_{pos}(\cdot \mid k,u,i^+)}
  \left[
    - \sum_{\ell \in \mathcal{M}}
      \log p_{\theta}\bigl(c_{i^+}^{\ell} \mid \tilde{\mathbf{c}}_{i^+}, \mathbf{s}_u, \mathbf{d}_{k}\bigr)
  \right],
\label{loss}
\end{equation}
\end{small}

\subsection{Inference Stage}
During inference, for each user $u$, we start from a fully masked SID
$\mathbf{x}_{(0)} = (\texttt{[MASK]}, \dots, \texttt{[MASK]})$ and iteratively denoise it conditioned on the user history $s_u$ and the initial difficulty vector $\mathbf{d}_{(L)}$, corresponding to the maximum mask count. Since recommendation requires a Top-$B$ set of items rather than a single sequence, we adopt a parallel beam-search decoding procedure. At each step, we (i) decide how many positions to update, (ii) select the specific positions according to model confidence, and (iii) perform beam search expansion on these positions. We next describe these three components in detail.

\subsubsection{Warm-up-based two-stage position scheduling strategy}
To efficiently decode SIDs, a straightforward idea is to update multiple positions in each iteration. However, since inference starts from a fully noised state, decoding multiple positions at the beginning is problematic. The model has not yet identified the key semantic dimensions, prediction confidence is generally low across positions, and incorrect tokens are likely to be filled in early. To address this, we adopt a warm‑up–based two-stage position scheduling strategy. We set a warm‑up step hyperparameter $R_{warm}<L$. At decoding step $r$, the number of positions to be updated in this step $m$ is:
\begin{small}
    \begin{equation}
        m=
\begin{cases}
1, & r < R_{\text{warm}},\\[2pt]
\ m_{par}, & r \ge R_{\text{warm}},m_{par}>1.
\end{cases}
    \end{equation}
\end{small}
In the first $R_{\text{warm}}$ steps, we decode only one position, always choosing the unfilled position with the highest confidence so that a few initial steps lock the strongest global semantic constraints. Note that position selection remains globally free rather than left-to-right. After this warm-up stage, we switch to the parallel stage. At each step, multiple highest-confidence positions $m_{par}$ are selected and denoised simultaneously, greatly reducing the decoding steps.

\subsubsection{Confidence-guided token position selection}
After determining the number of positions to decode $m$ at step $r$, we need to select the specific positions. For any current path $\mathbf{x}_{(r)} $, let its set of unfilled positions be $\mathcal{M}_{(r)}$. We feed $\mathbf{x}_{(r)}$, $s_u$ and $\mathbf{d_{(r)}}$ into the MDGR, and obtain the predicted distributions over the codebook for all unfilled positions $\mathbf{x}^\ell_{(r)}$:
\begin{equation}
p_{\theta}\big(\mathbf{x}^\ell_{(r)} = c \mid \mathbf{x}_{(r)}, \mathbf{s}_u, \mathbf{d}_{(r)}\big),
\quad \ell \in \mathcal{M}_{(r)},\ c \in \mathcal{V}_{\ell}.  
\end{equation}
To decide at which positions to actually generate tokens in this step, we first compute confidence scores over positions. For each unfilled position $\ell$, we take the probability of the most likely token at that position as its confidence:
\begin{equation}
\mathrm{conf}(\ell)
= \max_{c \in \mathcal{V}_{\ell}}
    p_{\theta}\big(\mathbf{x}^\ell_{(r)} = c \mid \mathbf{x}_{(r)}, \mathbf{s}_u, \mathbf{d}_{(r)}\big).
\end{equation}
Then, among all unfilled positions on this path, we select the $m$ positions with the highest confidence as the tokens to be decoded in this step for this path:
\begin{equation}
\mathcal{S}_{(r)} = \{\ell_1, \ldots, \ell_m\} \subset \mathcal{M}_{(r)},  
\end{equation}

\subsubsection{Beam Search-based candidate item generation}
After determining the position set $\mathcal{S}_{(r)}$ to be updated for this path at step $r$, we expand this path at these positions using a unified beam width $B$, and keep $B$ paths as candidate items. Before decoding, each beam $b \in \{1,\ldots,B\}$ maintains a SID sequence $\mathbf{x}_{(r),b} = (c_{(r),b}^{1},\ldots,c_{(r),b}^{L})$ (with some positions still masked) together with an accumulated log-score $v_{(r),b}$. Given $\mathcal{S}_{(r)}$, the decoder predicts a distribution over the corresponding codebook at each selected position $\ell \in \mathcal{S}_{(r)}$. For position $\ell$, we retain the Top-$B$ candidate tokens:
\begin{equation}
\{(\tilde{c}^{\ell,j}_{(r),b},\,\tilde{p}^{\ell,j}_{(r),b})\}_{j=1}^{B},
\end{equation}
where $\tilde{c}^{\ell,j}_{(r),b}$ denotes the $j$-th candidate token at position $\ell$ for beam $b$, and $\tilde{p}^{\ell,j}_{(r),b}$ is its predicted probability.

We then jointly expand the $m  $ positions. For every index tuple $(j_1,\dots,j_m)$ with $j_k \in \{1,\dots,B\}$, we construct a new candidate sequence $\mathbf{x}_{(r+1),(j_1,..,j_m)}$ by replacing $c^{\ell}_{(r),b}$ with $\tilde{c}^{\ell,j_k}_{(r),b}$ at $\ell$:

\begin{equation}
\mathbf{x}_{(r+1),(j_1,..,j_m)} =
\begin{cases}
\tilde{c}_{(r),b}^{\ell,j_k}, & \ell \in \mathcal{S}^{(r)}, k = 1,\ldots,m, \\[2pt]
c_{(r),b}^{\ell},   & \text{otherwise}.
\end{cases}
\end{equation}
Its log-score is updated as
\begin{equation}
v_{(r+1),(j_1,\dots,j_m)}
= v_{(r),b} + \sum_{k=1}^{m} \log \tilde{p}_{(r),b}^{\ell,j_k}. 
\end{equation}
Thus, each original beam $b$ is expanded into $B^m$ candidate beams at step $r$. Collecting all candidates across beams, we select the Top-$B$ according to $\tilde{v}$ to form the beam set for the next
step:
\begin{small}
\begin{equation}
\mathbf{x}_{(r+1)}
= \operatorname{Top}_B
   \Bigl\{
      \bigl(\mathbf{x}_{(r+1),(j_1,..,j_m)},
             v_{(r+1),(j_1,\dots,j_m)}\bigr)
   \Bigr\}_{(r+1),\,(j_1,\dots,j_m)}.
\end{equation}
\end{small}
At each diffusion step, only positions in $\mathcal{S}_{(r)}$ are updated, while the remaining positions stay masked and can be revised later. The procedure terminates when all positions are filled or a maximum number of steps is reached, yielding $B$ complete SIDs for each user, which are finally mapped back to items.

\begin{table}[t]
\centering
\caption{Asymptotic computational complexity of autoregressive GR, parallel GR, and MDGR in training and inference. }
\label{tab:complexity}
\resizebox{0.46\textwidth}{!}{
\begin{tabular}{lccc}
\toprule
\textbf{Stage}          & \textbf{Autoregressive}       & \textbf{Parallel}             & \textbf{MDGR}                    \\
\midrule
Training       & $O(H \cdot L^2 \cdot d)$  & $O(H \cdot L^2 \cdot d)$    & $O(H \cdot L^2 \cdot d)$    \\
Inference      & $O(B \cdot H \cdot L^3 \cdot d)$ & $O(B \cdot H \cdot L^2 \cdot d)$   & $O(R \cdot B \cdot H \cdot L^2 \cdot d)$    \\
\bottomrule
\end{tabular}}
\vspace{-10pt}
\end{table}

\subsection{Complexity Analysis}
\label{sec:Complexity_Analysis}
In this section, we analyze the training and inference complexity of diffusion-based GR as shown in Table \ref{tab:complexity}, and in the experiments we further study the trade-off between performance and efficiency from a numerical perspective  (Sec.~\ref{sec:trade-off}).

\begin{table*}[t]
\centering
\caption{Performance comparison on the Amazon Electronics, Amazon Books and Industrial datasets. "\textbf{Improv.}" shows the relative improvement (\%) over the base model. Best results are in \textbf{bold} and second-best are \underline{underlined}.}
\label{tab:amazon}
\resizebox{\textwidth}{!}{
\begin{tabular}{lcccccccccccc}
\toprule
\multirow{2}{*}{Methods} &
\multicolumn{4}{c}{\textbf{Amazon Electronics}} &
\multicolumn{4}{c}{\textbf{Amazon Books}} &
\multicolumn{4}{c}{\textbf{Industry}} \\
\cmidrule(lr){2-5} \cmidrule(lr){6-9} \cmidrule(lr){10-13}
& Recall@5 & NDCG@5 & Recall@10 & NDCG@10
& Recall@5 & NDCG@5 & Recall@10 & NDCG@10
& Recall@5 & NDCG@5 & Recall@10 & NDCG@10 \\
\midrule
\midrule
\multicolumn{13}{c}{\emph{Item ID-based Discriminative Recommendation}} \\
\midrule
SASRec       & 0.0321 & 0.0213 & 0.0490 & 0.0309
             & 0.0506 & 0.0297 & 0.0697 & 0.0391
             & 0.1413 & 0.0909 & 0.1695 & 0.1080 \\
BERT4Rec     & 0.0324 & 0.0214 & 0.0495 & 0.0311
             & 0.0509 & 0.0299 & 0.0699 & 0.0394
             & 0.1421 & 0.0911 & 0.1693 & 0.1079 \\
S$^3$Rec     & 0.0320 & 0.0211 & 0.0489 & 0.0307
             & 0.0514 & 0.0302 & 0.0704 & 0.0396
             & 0.1456 & 0.0934 & 0.1722 & 0.1101 \\
PinnerFormer & 0.0330 & 0.0220 & 0.0502 & 0.0316
             & 0.0525 & 0.0307 & 0.0716 & 0.0401
             & 0.1483 & 0.0950 & 0.1766 & 0.1132 \\
HeterRec     & 0.0336 & 0.0221 & 0.0510 & 0.0319
             & 0.0522 & 0.0306 & 0.0712 & 0.0398
             & 0.1592 & 0.1014 & 0.1890 & 0.1208 \\
DiffuRec     & 0.0334 & 0.0220 & 0.0507 & 0.0317
             & 0.0531 & 0.0311 & 0.0722 & 0.0403
             & 0.1586 & 0.1013 & 0.1889 & 0.1207 \\
VQ-Rec       & 0.0331 & 0.0217 & 0.0498 & 0.0312
             & 0.0517 & 0.0301 & 0.0687 & 0.0384
             & 0.1520 & 0.0975 & 0.1843 & 0.1177 \\
\midrule
\midrule
\multicolumn{13}{c}{\emph{Semantic ID-based Generative Recommendation}} \\
\midrule
TIGER        & 0.0348 & 0.0230 & 0.0524 & 0.0324
             & 0.0552 & 0.0315 & 0.0731 & 0.0408
             & 0.1620 & 0.1049 & 0.1943 & 0.1244 \\
Cobra       & \underline{0.0359} & \underline{0.0237} & \underline{0.0544} & 0.0338
             & 0.0576 & 0.0330 & \underline{0.0763} & 0.0426
             & 0.1659 & 0.1059 & 0.1996 & 0.1285 \\
RPG           & 0.0357 & 0.0236 & 0.0543 & \underline{0.0339}
             & \underline{0.0581} & \underline{0.0332} & 0.0762 & \underline{0.0427}
             & \underline{0.1679} & \underline{0.1076} & \underline{0.2018} & \underline{0.1296} \\
\midrule
\midrule
\multicolumn{13}{c}{\emph{Ours}} \\
\midrule
\textbf{MDGR} & \textbf{0.0394} & \textbf{0.0258} & \textbf{0.0583} & \textbf{0.0367}
              & \textbf{0.0631} & \textbf{0.0361} & \textbf{0.0826} & \textbf{0.0463}
              & \textbf{0.1856} & \textbf{0.1192} & \textbf{0.2210} & \textbf{0.1422} \\
\textbf{Improv}. & +9.75\% & +8.86\% & +7.17\% & +8.26\%
                 & +8.61\% & +8.73\% & +8.26\% & +8.43\%
                 & +10.45\% & +10.78\% & +9.54\% & +9.72\% \\
\bottomrule
\end{tabular}}
\end{table*}

\subsubsection{Complexity of Training}

During training, our model performs a single masked diffusion-style reconstruction per sample. Let SID length be $L$, decoder depth $H$, and hidden dimension $d$. A forward pass through the Transformer decoder has complexity $O(H \cdot L^2\cdot d)$, as in a standard Transformer. MDGR’s one-step denoising uses one forward and one backward pass, so the per-step complexity remains $O(H \cdot L^2\cdot d)$, comparable to an autoregressive Transformer of similar size. Curriculum noise scheduling and history-aware masking introduce only $O(L)$ extra operations (\textit{e.g.}, sampling noise levels and mask locations), which is negligible compared with $O(H \cdot L^2\cdot d)$. Thus, compared with autoregressive and parallel GRs of the same scale, our method maintains the same asymptotic training complexity while using difficulty-aware noise scheduling to provide more informative supervision at almost no extra cost.

\subsubsection{Complexity of Inference}
In autoregressive GR with beam search, the decoder must generate a length‑$L$ SID sequentially, deciding one position per step. Let the beam width be $B$. At each step, a forward pass over the $B$ candidate sequences dominates the cost, with complexity $O(B \cdot H \cdot L^2 \cdot d)$ from self‑attention and feed‑forward layers, while beam scoring and pruning are negligible. Because all $L$ positions are decoded in sequence, the overall inference complexity is $O(B \cdot H \cdot L^3 \cdot d)$. For standard parallel GR, the model predicts all $L$ positions in one shot for each decoding step, so the single-step complexity is $O(B \cdot H \cdot L^2 \cdot d)$. In contrast, our model can fill multiple groups of tokens in parallel at each step of parallel stage, and completes generation within $R=R_{warm}+\frac{L-R_{warm}}{m_{par}}$ denoising steps, with a time complexity of $O(R \cdot B \cdot H \cdot L^2 \cdot d)$.
Therefore, under the same model size and beam width, our parallel denoising reduces the decoding cost to approximately $\frac{R}{L}$ of that of autoregressive GR.

\section{Experiments}
To comprehensively evaluate MDGR, we design experiments to answer the following questions:
\begin{itemize}[noitemsep, topsep=0pt, leftmargin=*]
    \item \textbf{RQ1 Performance comparison:}  Does our method achieve better performance than existing discriminative models and generative recommenders?
    \item \textbf{RQ2 Efficiency and effectiveness trade-off: }Under different inference settings, how does the model balance recommendation quality and inference cost?
    \item \textbf{RQ3 Component contribution: } What is the role of each module in the MDGR?
    \item \textbf{RQ4 Hyperparameter sensitivity: }How do different hyperparameter choices affect recommendation performance?
    \item \textbf{RQ5 Online effectiveness: }How well does the model perform in online environments?
\end{itemize}

\subsection{Experimental Setup}
\subsubsection{Datasets and Evaluation Metrics.}
We evaluate our approach on two public datasets and one industrial dataset. The public datasets are two subsets of the Amazon Product Review corpus, Electronics and Books \cite{AmazonDataset}. 
Electronics contains about 250K users, 90K items, and 2.1M interactions, while Books contains about 110K users, 180K items, and 3.1M interactions. 
The industrial dataset consists of internal interaction logs from a major Southeast Asian e‑commerce platform. It contains over 1 billion user–item interactions from 18M users and 25M items, collected between May and December 2025, providing a realistic view of large‑scale user behavior in production. For each user, we record behavior sequences (including clicks and conversions) with an average length of 128 interactions. Each item is associated with rich multimodal content, including product images, titles, and textual descriptions.
For the public datasets, following prior work~\cite{tiger}, we remove users with fewer than five interactions, sort each user’s interactions chronologically, and cap the maximum sequence length at 20. We train with a sliding-window next-item prediction setup, where the model observes a prefix and predicts the subsequent item. For evaluation, we use the leave-one-out protocol~\cite{sasrec}: the most recent interaction is used for testing, the second most recent for validation, and the rest for training. 

We evaluate recommendation performance using two standard metrics: Recall@5/10 (R@5/10) and NDCG@5/10 (N@5/10) \cite{tiger,unisrec}.



\subsubsection{Baseline Models.}
To ensure a comprehensive evaluation, we compare our method with both traditional ID-based approaches and the latest SID-based GRs. \textbf{ID-based methods:} \textbf{(1) SASRec} \cite{sasrec} uses a unidirectional self-attention network for sequence modeling. \textbf{(2) HeterRec} \cite{deng2025heterrec} employs a dual-tower hierarchical Transformer to model multi-modal item features, together with a multi-step list-wise prediction loss. \textbf{(3) Pinnerformer} \cite{pancha2022pinnerformer} uses a causally masked Transformer to model long-term user behaviors. \textbf{(4) S$^3$Rec} \cite{zhou2020s3} enhances representations via self-supervised learning. \textbf{(5) Bert4Rec} \cite{bert4rec} adopts a bidirectional Transformer to capture contextual user interests. \textbf{(6) DiffuRec} \cite{li2023diffurec} introduces diffusion models into sequential recommendation, replacing fixed item embeddings with distributional representations that can model uncertainty. \textbf{(7) VQ-Rec} \cite{hou2023learning} builds an item codebook with VQ-VAE and obtains item representations via pooling.
\textbf{SID-based GRs:} \textbf{(1) TIGER} \cite{tiger} constructs a codebook with RQ-VAE \cite{lee2022autoregressive} and autoregressively generates tokens using an encoder–decoder architecture. \textbf{(2) RPG} \cite{rpg} proposes a parallel generation framework for long SIDs, leveraging multi-token prediction and graph constraints for parallel decoding. \textbf{(3) Cobra} \cite{cobra} integrates sparse SIDs with dense vectors to alleviate the information loss caused by directly using semantic tokens in generative recommendation. 

\subsubsection{Implementation Details.} Our experiments are conducted on a distributed PyTorch platform \cite{paszke2019pytorch} with 2 parameter servers and 10 workers, each equipped with a single Nvidia A100 GPU. \textbf{(1) Codebook construction:} we use a pretrained Qwen3-8B \cite{bai2023qwen} model to encode the content modality of items and obtain their representation vectors. 
For the parallel codebook, we use 8 separate codebooks, each of size 300 with 256-dimensional code vectors, resulting in an 8-token SID for each item.
\textbf{(2) Training Stage:} we adopt a standard encoder–decoder architecture \cite{vaswani2017attention}. On the encoder side, a 6-layer Transformer with hidden size 256 and 8 attention heads is used to model click sequences. On the decoder side, we use bidirectional attention. The parameter $\gamma$ is set to 2.0 and the training objective is the standard cross-entropy loss as defined in the Eq.(\ref{loss}). \textbf{(3) Inference Stage:} the initial SID is set to an all \texttt{[MASK]} sequence. We set $R_{warm}$ to 4, i.e., in the first 4 steps we decode only one position per step, always choosing the position with the highest confidence. The subsequent 4 steps switch to parallel mode with $m_{par}$ set to 2. We set the beam width to $B=50$ for all experiments. Decoding stops once all positions are filled. For the generated SIDs, we map each SID back to the concrete item set using the offline-constructed codebook index \cite{xing2025reg4rec}.

\subsection{Overall Performance (RQ1)}
We compare MDGR with existing ID-based discriminative models and GRs on 2 public datasets and a real-world industrial dataset. The results are reported in Table \ref{tab:amazon}, from which we observe that:
\begin{itemize}[noitemsep, topsep=0pt, leftmargin=*]
    \item Our method achieves the best performance on all datasets and metrics. On the industrial dataset, for example, MDGR improves over the strongest baseline by 10.45\%/10.78\% in R@5/N@5 and 9.54\%/9.72\% in R@10/N@10, indicating that the diffusion-based GR framework can more accurately decode users’ dynamic and heterogeneous interests.
    \item Compared with existing GRs, MDGR achieves 7.17\%–10.78\% relative gains across all metrics. We do not substantially modify the way codebooks are constructed; instead, we mainly redesign the training supervision and decoding mechanism. This suggests that the discrete diffusion learning process is better suited for GR and enforces stronger global consistency across multiple semantic dimensions than autoregressive decoding. By reformulating recommendation as a discrete mask denoising task and introducing curriculum-style noise scheduling and history-aware masking, MDGR better captures user preferences under different missing patterns, leading to the observed improvements.

\end{itemize}

\begin{table}[t]
\centering
\caption{Impact of $R_{warm}$ and $m_{par}$ on performance and efficiency, where Recall@10 represents recommendation performance and QPS represents inference efficiency.}
\label{tab:warmup_m}
\resizebox{0.92\linewidth}{!}{
\begin{tabular}{cccccc}
\toprule
$\mathbf{R_{warm}}$ & $\mathbf{m_{par}}$& \textbf{Total step} & \textbf{Recall@10} & \textbf{QPS}  & \textbf{QPS uplift} \\
\midrule
\midrule
\multicolumn{6}{c}{\emph{Base Setting}} \\
\midrule
0 & 1 & 8 & 0.2219 & 2.212 &  --      \\
\midrule
\midrule
\multicolumn{6}{c}{\emph{Impact of $R_{warm}$}} \\
\midrule
0 & 2 & 4 & 0.2127 & 3.015 & +36.30\%  \\
2 & 2 & 5 & 0.2199 & 2.741 & +23.92\%  \\
4 & 2 & 6 & \textbf{0.2210} &\textbf{ 2.527 }& \textbf{+14.24\% } \\
6 & 2 & 7 & 0.2213 & 2.365 & +6.92\%   \\
\midrule
\midrule
\multicolumn{6}{c}{\emph{Impact of $m_{par}$}} \\
\midrule
4 & 3 & 6 & 0.2205 & 2.526 & +14.19\%  \\
4 & 4 & 5 & 0.2204 & 2.739 & +23.82\%  \\
\bottomrule
\end{tabular}
}
\vspace{-10pt}
\end{table}

\subsection{The trade-off between efficiency and performance }
\label{sec:trade-off}
In this section, we focus on how key hyperparameters at inference time affect the trade-off between efficiency and performance in the industrial dataset, specifically: (i) the number of warm-up steps with single-position decoding $R_{warm}$; (ii) the number of positions decoded per step in the parallel stage $m_{par}$.

\begin{table}[t]
\centering
\small
\caption{Ablation study of MDGR on the industrial dataset.}
\label{tab:ablation}
\resizebox{0.95\linewidth}{!}{
\begin{tabular}{lcccc}
\toprule
\textbf{Setting} & \textbf{Recall@5} & \textbf{Improv.} & \textbf{NDCG@5} &  \textbf{Improv.} \\
\midrule
\midrule
\multicolumn{5}{c}{\emph{Base Setting}} \\
\midrule
MDGR                   & \textbf{0.1856} & \textbf{0}        & \textbf{0.1192} & \textbf{0    }    \\
\midrule
\midrule
\multicolumn{5}{c}{\emph{Impact of Codebook}} \\
\midrule
RQ-VAE          & 0.1829 & -1.46\% & 0.1175 & -1.43\% \\
RQ-KMeans       & 0.1830 & -1.40\% & 0.1176 & -1.34\%\\
\midrule
\midrule
\multicolumn{5}{c}{\emph{Ablation of Training Stage}} \\
\midrule
Random quantity& 0.1831 & -1.35\% & 0.1179 & -1.09\%\\
Random positions    & 0.1822 & -1.83\% & 0.1168 & -2.01\% \\
Vanilla mask       & 0.1792 & -3.45\% & 0.1154 & -3.19\% \\
w/o $\mathbf{d}_k$ & 0.1837 & -1.02\% & 0.1179 & -1.09\% \\
\midrule
\midrule
\multicolumn{5}{c}{\emph{Ablation of Inference Stage}} \\
\midrule
w/o confidence    & 0.1820 & -1.94\% & 0.1170 & -1.85\% \\
\bottomrule
\end{tabular}
}
\vspace{-15pt}
\end{table}

\subsubsection{The impact of $R_{warm}$}
We first fix $m_{par}=2$ and vary $R_{warm}$ to examine the changes in queries per second (QPS, reflecting inference speed) and Recall@10 (reflecting model performance). We set $R_{warm} \in \{0,2, 4,6,8\}$, where $R_{warm}=0$ means no warm-up, i.e., decoding 2 tokens in every step from the beginning. The results are shown in Table \ref{tab:warmup_m}, where the first row (baseline) corresponds to the MDGR without parallel decoding acceleration, i.e., decoding only one position at each step. Under the no–warm-up setting, inference becomes much faster, with QPS improved by 36.30\%, indicating that the parallel strategy can substantially accelerate decoding compared with step-wise serial decoding. However, recommendation performance drops 4.15\% in this case. As we increase $R_{warm}$, the performance gradually recovers and eventually approaches or even matches that of full single-position decoding. In addition, the performance is stable when $R_{warm} \in \{4,6\}$, suggesting that using only a few warm-up steps to stabilize key semantic pivots is sufficient to enable accurate parallel decoding and mitigate error amplification.

\subsubsection{The impact of $m_{par}$}
We further study the impact of the number of positions decoded per step in the parallel stage $m_{par}$. To this end, we fix $R_{warm}=4$ and vary $m_{par} \in \{1,2,3,4\}$, recording the changes in QPS and R@10. As $m_{par}$ increases, QPS consistently improves, while recommendation performance degrades noticeably. When $m_{par}=2$, the combinatorial space expanded at each step is smaller, allowing beam search to better cover high-probability paths under a fixed beam width, thus Recall@10 is almost unchanged compared with full single-position decoding, with only minor fluctuations. As $m_{par}$ increases, QPS keeps increasing. However, the number of combinations $B^{m_{par}}$ expanded at each step grows sharply. Under a fixed beam width, the search space becomes much more compressed, causing some high-quality paths to be pruned, which in turn reduces Recall@10.

In summary, we find that a suitable choice of $R_{warm}$=4 and $m_{par}=2$ can usually strike a better balance between inference speed and recommendation performance. We adopt this setting as the default decoding configuration in subsequent experiments.

\subsection{Ablation Study (RQ3)}
\label{sec:Ablation}
In this subsection, we conduct ablation studies to evaluate the contribution of each component on the industrial dataset from three perspectives: codebook, training, and inference.
\textbf{(1) Codebook structure.} We replace our parallel codebook with two typical residual quantization schemes, RQ-VAE \cite{lee2022autoregressive} and RQ-KMeans \cite{zhou2025onerec}, and compare their differences in semantic representation capability and downstream recommendation performance. 
\textbf{(2) Training strategy.} To study the effect of curriculum-style noise scheduling and history-aware mask allocation, we construct the following variants:
\begin{itemize}[noitemsep, topsep=0pt, leftmargin=*]
    \item Random noise quantity: we keep the history-aware allocation of mask positions, but no longer use curriculum scheduling along the temporal dimension; instead, the number of masked tokens is sampled from a uniform distribution.
    \item Random noise positions: we keep only the total number of masks given by curriculum scheduling, but at the sample level we uniformly sample mask positions over all locations.
    \item Vanilla mask: both the number and positions of masks are sampled from a uniform distribution, corresponding to a vanilla masked diffusion (MDM-style) scheme without adapting the corruption pattern to user-interest structure or task difficulty.
\end{itemize}
In addition, while keeping curriculum scheduling and history-aware allocation, we remove the difficulty-aware vector $\mathbf{d}_{k}$ to assess its additional gain. \textbf{(3) Inference. }Section~\ref{sec:trade-off} has analyzed the effect of warm-up and parallel filling. Here, we further replace confidence-based position selection with random selection from unfilled positions, to examine the role of confidence-guided position choice in the parallel denoising process.
   
\begin{figure}[t]
\centering
\begin{minipage}[b]{0.48\columnwidth}
\centering
\includegraphics[width=\linewidth]{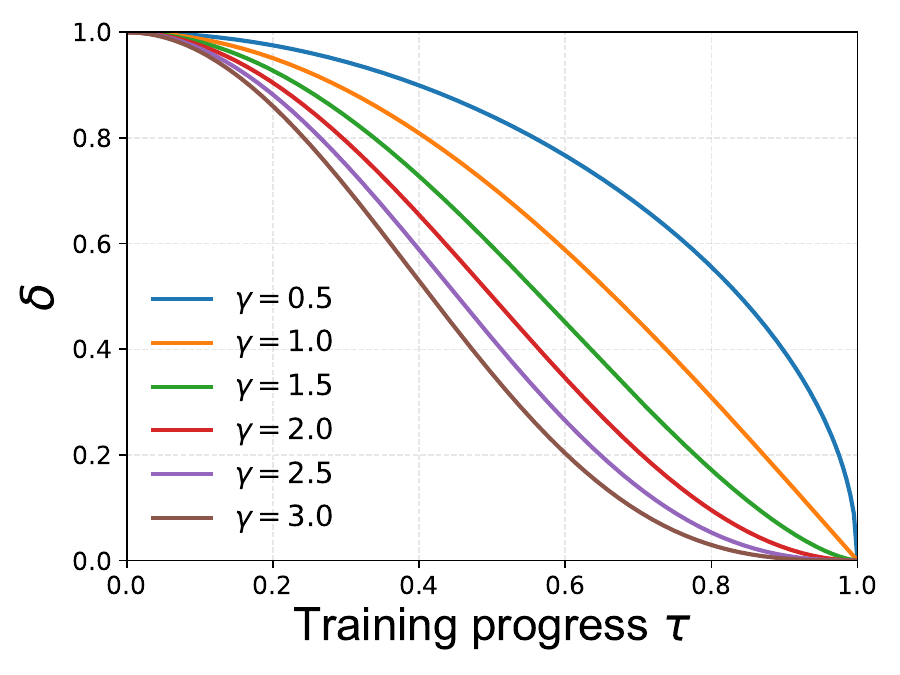}
\end{minipage}\hfill
\begin{minipage}[b]{0.48\columnwidth}
\centering
\includegraphics[width=\linewidth]{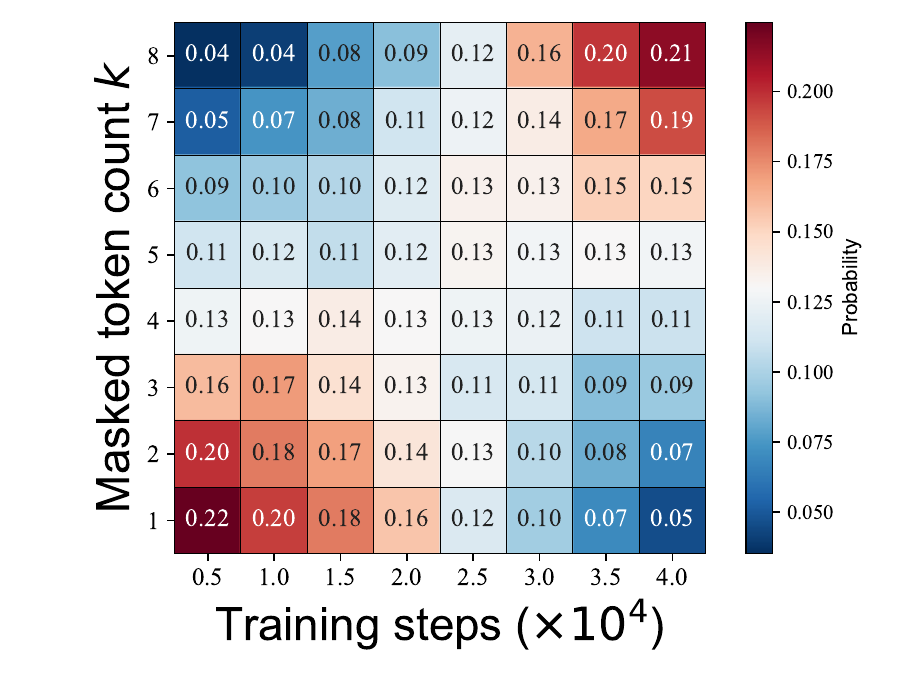}
\end{minipage}
\caption{(a) Effect of the $\gamma$ on the global difficulty schedule. (left). (b) Empirical distribution of masked‑token counts $k$ over training steps when $\gamma=2$ (right).}
\label{fig:two-minipage}

\vspace{-10pt}
\end{figure}

\subsubsection{Codebook structure.}
According to Table \ref{tab:ablation}, the parallel codebook consistently outperforms both alternatives on all metrics. This is because the parallel codebook does not impose strict dependencies in its structure. Each sub-codebook corresponds to a relatively independent semantic dimension, allowing the decoder to perform denoising predictions at arbitrary positions and in arbitrary orders, which better matches the heterogeneous interest structure of users. In contrast, residual quantization ties lower-level tokens to the residuals of upper levels, making it prone to error accumulation under the diffusion process.

\subsubsection{Training stage}
(1) Removing the global curriculum noise scheduling (Random quantity) leads to a 1.35\%/1.09\% drop in Recall@5/NDCG@5, showing that gradually increasing noise difficulty is more effective than sampling mask counts uniformly from the beginning. (2) Randomly assigning mask positions (Random positions) further degrades performance by 1.83\%/2.01\%, and fully random masking (Vanilla mask), i.e., standard MDM-style uniform masking over both mask counts and positions, yields the largest degradation (3.45\%/3.19\%). This pattern indicates that naive discrete diffusion with uniform masking is less capable of capturing users’ heterogeneous interests and maintaining multi-attribute consistency.
(3) Dropping the difficulty-aware vector leads to a small decline, suggesting that explicitly encoding corruption difficulty helps stabilize optimization and slightly improves final accuracy.

\subsubsection{Inference stage}
According to Table \ref{tab:ablation}, randomly selecting decoding positions leads to a clear performance drop, showing that the choice of positions in each parallel denoising step affects the final result. Updating the most confident positions first helps the model quickly fix key semantic anchors and provides reliable context for later decoding, whereas random updates may introduce early errors at uncertain positions that beam search can hardly correct, reducing the overall consistency of the generated SID.


\subsection{Analysis Experiments (RQ4)}
This section studies how the exponent $\gamma$ shapes the global noise schedule and influences recommendation quality. We visualize the resulting difficulty and masking patterns over training, and then conduct an ablation on different $\gamma$ values to measure their effect on Recall and NDCG in the industrial dataset.

\subsubsection{Visualization}
We first visualize how $\gamma$ reshapes the global difficulty schedule in Eq.~(\ref{eq:stretched_difficulty}). Figure \ref{fig:two-minipage} (a) plots $\delta$ as a function of training progress for different $\gamma$. Larger $\gamma$ flattens the curve in early stages and makes difficulty grow faster later, yielding a more conservative curriculum. To understand the resulting masking behavior, Figure \ref{fig:two-minipage} (b) shows, for $\gamma=2$, a heatmap of the empirical distribution of the masked‑token count $k\in\{1,...,8\}$ at training steps. As training proceeds, probability mass gradually shifts from small to large $k$, confirming that the model is exposed to increasingly harder SID reconstruction tasks (with more heavily masked tokens).

\begin{figure}[t]
\centering
\begin{minipage}[b]{0.48\columnwidth}
\centering
\includegraphics[width=\linewidth]{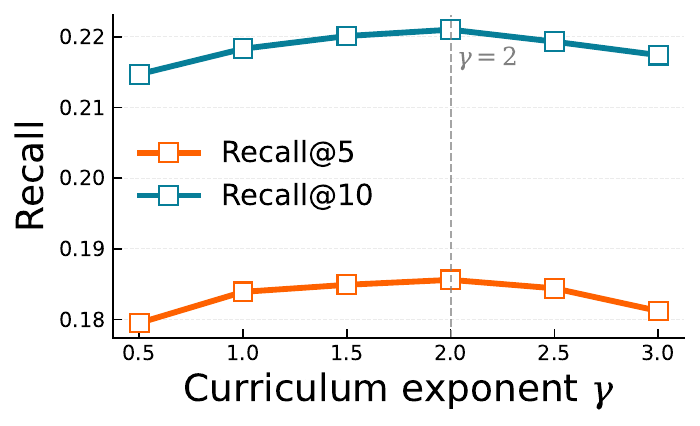}
\end{minipage}\hfill
\begin{minipage}[b]{0.48\columnwidth}
\centering
\includegraphics[width=\linewidth]{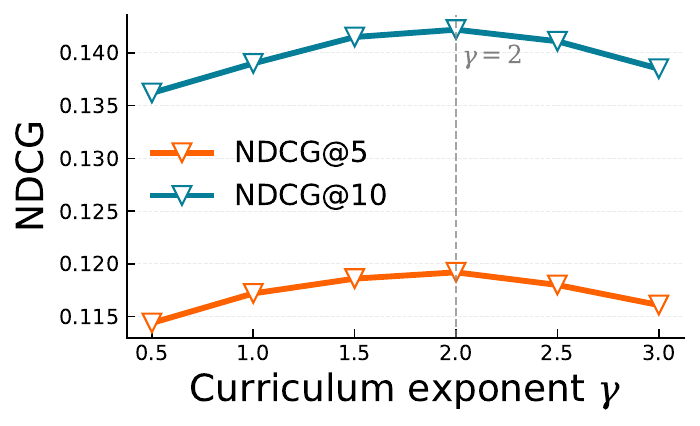}
\end{minipage}
\caption{(a) Effect of the curriculum exponent $\gamma$ on Recall (left). (b) Effect of $\gamma$ on NDCG (right).}
\label{fig:gamma}
\vspace{-10pt}
\end{figure}
\subsubsection{Impact of $\gamma$}

We further study how $\gamma$ affects performance. As shown in Figure \ref{fig:gamma}, performance first improves and then degrades as $\gamma$ increases. A too small $\gamma$ (0.5–1.0) makes the curriculum overly aggressive, yielding lower Recall and NDCG, while a too large $\gamma$(2.5–3.0) slows down exposure to hard samples and again harms performance. $\gamma=2$ achieves the best results on all metrics, and is therefore adopted as the default setting.

\subsection{Online Experiments (RQ5)}
We further evaluate MDGR through an online A/B test on the advertising recommendation platform of a leading e-commerce company in Southeast Asia, conducted from Jan 5 to 12, 2026. The baseline system in production is TIGER \cite{tiger}, which serves as the control group, while the experimental group replaces it with our proposed MDGR. Each group contains 20\% of users sampled uniformly at random. Compared with the control, MDGR delivers a \textbf{1.20\%} lift in \textbf{advertising revenue}, a \textbf{3.69\%} increase in \textbf{gross merchandise volume (GMV)}, and a \textbf{2.36\%} improvement in \textbf{click-through rate (CTR)}, all statistically significant under a two-sided test ($p < 0.05$). These online gains demonstrate the practical effectiveness of MDGR in a large-scale industrial environment.



\section{Conclusion}
In this work, we propose MDGR, a masked diffusion generative recommendation framework that departs from the traditional autoregressive paradigm. We identify three key limitations of autoregressive methods: difficulty in modeling global dependencies among multi-dimensional semantics, inability to adapt to heterogeneous user interests, and low inference efficiency.
MDGR redesigns GR from three aspects: codebook, training, and inference. It uses a parallel codebook to quantize item embeddings into multiple semantic subspaces, enabling multi-position parallel generation. In training, MDGR introduces a discrete diffusion noise schedule along temporal and sample dimensions. In inference, it adopts a warm‑up–based two-stage parallel decoding scheme for efficient generation.
Experiments on multiple datasets show that MDGR outperforms both discriminative and generative baselines, suggesting that masked diffusion is a competitive new paradigm for GR. In future work, we will further improve diffusion-based GR with better noise sampling and interest-aligned decoding strategies.

\bibliographystyle{ACM-Reference-Format}
\bibliography{main}

@article{wang2021survey,
  title={A survey on session-based recommender systems},
  author={Wang, Shoujin and Cao, Longbing and Wang, Yan and Sheng, Quan Z and Orgun, Mehmet A and Lian, Defu},
  journal={ACM Computing Surveys (CSUR)},
  volume={54},
  number={7},
  pages={1--38},
  year={2021},
  publisher={ACM New York, NY, USA}
}

@article{wang2024rethinking,
  title={Rethinking large language model architectures for sequential recommendations},
  author={Wang, Hanbing and Liu, Xiaorui and Fan, Wenqi and Zhao, Xiangyu and Kini, Venkataramana and Yadav, Devendra and Wang, Fei and Wen, Zhen and Tang, Jiliang and Liu, Hui},
  journal={arXiv preprint arXiv:2402.09543},
  year={2024}
}

@inproceedings{lin2024enhancing,
  title={Enhancing Relevance of Embedding-based Retrieval at Walmart},
  author={Lin, Juexin and Yadav, Sachin and Liu, Feng and Rossi, Nicholas and Suram, Praveen R and Chembolu, Satya and Chandran, Prijith and Mohapatra, Hrushikesh and Lee, Tony and Magnani, Alessandro and others},
  booktitle={Proceedings of the 33rd ACM International Conference on Information and Knowledge Management},
  pages={4694--4701},
  year={2024}
}

@inproceedings{wang2025home,
  title={Home: Hierarchy of multi-gate experts for multi-task learning at kuaishou},
  author={Wang, Xu and Cao, Jiangxia and Fu, Zhiyi and Gai, Kun and Zhou, Guorui},
  booktitle={Proceedings of the 31st ACM SIGKDD Conference on Knowledge Discovery and Data Mining V. 1},
  pages={2638--2647},
  year={2025}
}

@article{gray1984vector,
  title={Vector quantization},
  author={Gray, Robert},
  journal={IEEE Assp Magazine},
  volume={1},
  number={2},
  pages={4--29},
  year={1984},
  publisher={IEEE}
}

@article{mu2025synergistic,
  title={Synergistic Integration and Discrepancy Resolution of Contextualized Knowledge for Personalized Recommendation},
  author={Mu, Lingyu and Deng, Hao and Xing, Haibo and Lin, Kaican and Zhu, Zhitong and Zhang, Yu and Zeng, Xiaoyi and Liu, Zhengxiao and Lin, Zheng and Hu, Jinxin},
  journal={arXiv preprint arXiv:2510.14257},
  year={2025}
}

@article{zhou2025onerec,
  title={OneRec Technical Report},
  author={Zhou, Guorui and Deng, Jiaxin and Zhang, Jinghao and Cai, Kuo and Ren, Lejian and Luo, Qiang and Wang, Qianqian and Hu, Qigen and Huang, Rui and Wang, Shiyao and others},
  journal={arXiv preprint arXiv:2506.13695},
  year={2025}
}

@article{rpg,
  title={Generating Long Semantic IDs in Parallel for Recommendation},
  author={Hou, Yupeng and Li, Jiacheng and Shin, Ashley and Jeon, Jinsung and Santhanam, Abhishek and Shao, Wei and Hassani, Kaveh and Yao, Ning and McAuley, Julian},
  journal={arXiv preprint arXiv:2506.05781},
  year={2025}
}

@inproceedings{wang2024learnable,
  title={Learnable item tokenization for generative recommendation},
  author={Wang, Wenjie and Bao, Honghui and Lin, Xinyu and Zhang, Jizhi and Li, Yongqi and Feng, Fuli and Ng, See-Kiong and Chua, Tat-Seng},
  booktitle={Proceedings of the 33rd ACM International Conference on Information and Knowledge Management},
  pages={2400--2409},
  year={2024}
}

@article{xing2025reg4rec,
  title={Reg4rec: Reasoning-enhanced generative model for large-scale recommendation systems},
  author={Xing, Haibo and Deng, Hao and Mao, Yucheng and Hu, Jinxin and Xu, Yi and Zhang, Hao and Wang, Jiahao and Wang, Shizhun and Zhang, Yu and Zeng, Xiaoyi and others},
  journal={arXiv preprint arXiv:2508.15308},
  year={2025}
}

@article{bai2023qwen,
  title={Qwen technical report},
  author={Bai, Jinze and Bai, Shuai and Chu, Yunfei and Cui, Zeyu and Dang, Kai and Deng, Xiaodong and Fan, Yang and Ge, Wenbin and Han, Yu and Huang, Fei and others},
  journal={arXiv preprint arXiv:2309.16609},
  year={2023}
}

@article{wang2023generative,
  title={Generative recommendation: Towards next-generation recommender paradigm},
  author={Wang, Wenjie and Lin, Xinyu and Feng, Fuli and He, Xiangnan and Chua, Tat-Seng},
  journal={arXiv preprint arXiv:2304.03516},
  year={2023}
}

@inproceedings{zhang2024generative,
  title={On generative agents in recommendation},
  author={Zhang, An and Chen, Yuxin and Sheng, Leheng and Wang, Xiang and Chua, Tat-Seng},
  booktitle={Proceedings of the 47th international ACM SIGIR conference on research and development in Information Retrieval},
  pages={1807--1817},
  year={2024}
}

@article{vaswani2017attention,
  title={Attention is all you need},
  author={Vaswani, Ashish and Shazeer, Noam and Parmar, Niki and Uszkoreit, Jakob and Jones, Llion and Gomez, Aidan N and Kaiser, {\L}ukasz and Polosukhin, Illia},
  journal={Advances in neural information processing systems},
  volume={30},
  year={2017}
}

@inproceedings{deng2025csmf,
  title={CSMF: Cascaded Selective Mask Fine-Tuning for Multi-Objective Embedding-Based Retrieval},
  author={Deng, Hao and Xing, Haibo and Matsuyama, Kanefumi and Zhang, Moyu and Hu, Jinxin and Wen, Hong and Zhang, Yu and Zeng, Xiaoyi and Zhang, Jing},
  booktitle={Proceedings of the 48th International ACM SIGIR Conference on Research and Development in Information Retrieval},
  pages={2122--2131},
  year={2025}
}

@article{yao2016things,
  title={Things of interest recommendation by leveraging heterogeneous relations in the internet of things},
  author={Yao, Lina and Sheng, Quan Z and Ngu, Anne HH and Li, Xue},
  journal={ACM Transactions on Internet Technology (TOIT)},
  volume={16},
  number={2},
  pages={1--25},
  year={2016},
  publisher={ACM New York, NY, USA}
}

@article{zhang2016modeling,
  title={Modeling the heterogeneous duration of user interest in time-dependent recommendation: A hidden semi-Markov approach},
  author={Zhang, Haidong and Ni, Wancheng and Li, Xin and Yang, Yiping},
  journal={IEEE Transactions on Systems, Man, and Cybernetics: Systems},
  volume={48},
  number={2},
  pages={177--194},
  year={2016},
  publisher={IEEE}
}

@inproceedings{lou2023reflected,
  title={Reflected diffusion models},
  author={Lou, Aaron and Ermon, Stefano},
  booktitle={International Conference on Machine Learning},
  pages={22675--22701},
  year={2023},
  organization={PMLR}
}

@article{graves2023bayesian,
  title={Bayesian flow networks},
  author={Graves, Alex and Srivastava, Rupesh Kumar and Atkinson, Timothy and Gomez, Faustino},
  journal={arXiv preprint arXiv:2308.07037},
  year={2023}
}

@inproceedings{lin2023text,
  title={Text generation with diffusion language models: A pre-training approach with continuous paragraph denoise},
  author={Lin, Zhenghao and Gong, Yeyun and Shen, Yelong and Wu, Tong and Fan, Zhihao and Lin, Chen and Duan, Nan and Chen, Weizhu},
  booktitle={International Conference on Machine Learning},
  pages={21051--21064},
  year={2023},
  organization={PMLR}
}

@article{xue2024unifying,
  title={Unifying bayesian flow networks and diffusion models through stochastic differential equations},
  author={Xue, Kaiwen and Zhou, Yuhao and Nie, Shen and Min, Xu and Zhang, Xiaolu and Zhou, Jun and Li, Chongxuan},
  journal={arXiv preprint arXiv:2404.15766},
  year={2024}
}

@article{zhang2025target,
  title={Target concrete score matching: A holistic framework for discrete diffusion},
  author={Zhang, Ruixiang and Zhai, Shuangfei and Zhang, Yizhe and Thornton, James and Ou, Zijing and Susskind, Joshua and Jaitly, Navdeep},
  journal={arXiv preprint arXiv:2504.16431},
  year={2025}
}

@article{austin2021structured,
  title={Structured denoising diffusion models in discrete state-spaces},
  author={Austin, Jacob and Johnson, Daniel D and Ho, Jonathan and Tarlow, Daniel and Van Den Berg, Rianne},
  journal={Advances in neural information processing systems},
  volume={34},
  pages={17981--17993},
  year={2021}
}

@article{lou2023discrete,
  title={Discrete diffusion language modeling by estimating the ratios of the data distribution},
  author={Lou, Aaron and Meng, Chenlin and Ermon, Stefano},
  year={2023}
}

@article{shi2024simplified,
  title={Simplified and generalized masked diffusion for discrete data},
  author={Shi, Jiaxin and Han, Kehang and Wang, Zhe and Doucet, Arnaud and Titsias, Michalis},
  journal={Advances in neural information processing systems},
  volume={37},
  pages={103131--103167},
  year={2024}
}

@article{sahoo2024simple,
  title={Simple and effective masked diffusion language models},
  author={Sahoo, Subham and Arriola, Marianne and Schiff, Yair and Gokaslan, Aaron and Marroquin, Edgar and Chiu, Justin and Rush, Alexander and Kuleshov, Volodymyr},
  journal={Advances in Neural Information Processing Systems},
  volume={37},
  pages={130136--130184},
  year={2024}
}

@article{ou2024your,
  title={Your absorbing discrete diffusion secretly models the conditional distributions of clean data},
  author={Ou, Jingyang and Nie, Shen and Xue, Kaiwen and Zhu, Fengqi and Sun, Jiacheng and Li, Zhenguo and Li, Chongxuan},
  journal={arXiv preprint arXiv:2406.03736},
  year={2024}
}

@article{nie2025large,
  title={Large language diffusion models},
  author={Nie, Shen and Zhu, Fengqi and You, Zebin and Zhang, Xiaolu and Ou, Jingyang and Hu, Jun and Zhou, Jun and Lin, Yankai and Wen, Ji-Rong and Li, Chongxuan},
  journal={arXiv preprint arXiv:2502.09992},
  year={2025}
}

@article{zhu2025llada,
  title={LLaDA 1.5: Variance-Reduced Preference Optimization for Large Language Diffusion Models},
  author={Zhu, Fengqi and Wang, Rongzhen and Nie, Shen and Zhang, Xiaolu and Wu, Chunwei and Hu, Jun and Zhou, Jun and Chen, Jianfei and Lin, Yankai and Wen, Ji-Rong and others},
  journal={arXiv preprint arXiv:2505.19223},
  year={2025}
}

@inproceedings{chang2022maskgit,
  title={Maskgit: Masked generative image transformer},
  author={Chang, Huiwen and Zhang, Han and Jiang, Lu and Liu, Ce and Freeman, William T},
  booktitle={Proceedings of the IEEE/CVF conference on computer vision and pattern recognition},
  pages={11315--11325},
  year={2022}
}

@article{chang2023muse,
  title={Muse: Text-to-image generation via masked generative transformers},
  author={Chang, Huiwen and Zhang, Han and Barber, Jarred and Maschinot, AJ and Lezama, Jose and Jiang, Lu and Yang, Ming-Hsuan and Murphy, Kevin and Freeman, William T and Rubinstein, Michael and others},
  journal={arXiv preprint arXiv:2301.00704},
  year={2023}
}

@article{you2025effective,
  title={Effective and efficient masked image generation models},
  author={You, Zebin and Ou, Jingyang and Zhang, Xiaolu and Hu, Jun and Zhou, Jun and Li, Chongxuan},
  journal={arXiv preprint arXiv:2503.07197},
  year={2025}
}

@article{tiger,
  title={Recommender systems with generative retrieval},
  author={Rajput, Shashank and Mehta, Nikhil and Singh, Anima and Hulikal Keshavan, Raghunandan and Vu, Trung and Heldt, Lukasz and Hong, Lichan and Tay, Yi and Tran, Vinh and Samost, Jonah and others},
  journal={Advances in Neural Information Processing Systems},
  volume={36},
  pages={10299--10315},
  year={2023}
}

@article{cobra,
  title={Sparse meets dense: Unified generative recommendations with cascaded sparse-dense representations},
  author={Yang, Yuhao and Ji, Zhi and Li, Zhaopeng and Li, Yi and Mo, Zhonglin and Ding, Yue and Chen, Kai and Zhang, Zijian and Li, Jie and Li, Shuanglong and others},
  journal={arXiv preprint arXiv:2503.02453},
  year={2025}
}

@article{ge2013optimized,
  title={Optimized product quantization},
  author={Ge, Tiezheng and He, Kaiming and Ke, Qifa and Sun, Jian},
  journal={IEEE transactions on pattern analysis and machine intelligence},
  volume={36},
  number={4},
  pages={744--755},
  year={2013},
  publisher={IEEE}
}

@inproceedings{hou2023learning,
  title={Learning vector-quantized item representation for transferable sequential recommenders},
  author={Hou, Yupeng and He, Zhankui and McAuley, Julian and Zhao, Wayne Xin},
  booktitle={Proceedings of the ACM Web Conference 2023},
  pages={1162--1171},
  year={2023}
}

@inproceedings{sasrec,
  title={Self-attentive sequential recommendation},
  author={Kang, Wang-Cheng and McAuley, Julian},
  booktitle={2018 IEEE international conference on data mining (ICDM)},
  pages={197--206},
  year={2018},
  organization={IEEE}
}

@inproceedings{bert4rec,
  title={BERT4Rec: Sequential recommendation with bidirectional encoder representations from transformer},
  author={Sun, Fei and Liu, Jun and Wu, Jian and Pei, Changhua and Lin, Xiao and Ou, Wenwu and Jiang, Peng},
  booktitle={Proceedings of the 28th ACM international conference on information and knowledge management},
  pages={1441--1450},
  year={2019}
}

@inproceedings{pancha2022pinnerformer,
  title={Pinnerformer: Sequence modeling for user representation at pinterest},
  author={Pancha, Nikil and Zhai, Andrew and Leskovec, Jure and Rosenberg, Charles},
  booktitle={Proceedings of the 28th ACM SIGKDD conference on knowledge discovery and data mining},
  pages={3702--3712},
  year={2022}
}

@inproceedings{zhou2020s3,
  title={S3-rec: Self-supervised learning for sequential recommendation with mutual information maximization},
  author={Zhou, Kun and Wang, Hui and Zhao, Wayne Xin and Zhu, Yutao and Wang, Sirui and Zhang, Fuzheng and Wang, Zhongyuan and Wen, Ji-Rong},
  booktitle={Proceedings of the 29th ACM international conference on information \& knowledge management},
  pages={1893--1902},
  year={2020}
}

@article{li2023diffurec,
  title={Diffurec: A diffusion model for sequential recommendation},
  author={Li, Zihao and Sun, Aixin and Li, Chenliang},
  journal={ACM Transactions on Information Systems},
  volume={42},
  number={3},
  pages={1--28},
  year={2023},
  publisher={ACM New York, NY}
}

@article{paszke2019pytorch,
  title={Pytorch: An imperative style, high-performance deep learning library},
  author={Paszke, Adam and Gross, Sam and Massa, Francisco and Lerer, Adam and Bradbury, James and Chanan, Gregory and Killeen, Trevor and Lin, Zeming and Gimelshein, Natalia and Antiga, Luca and others},
  journal={Advances in neural information processing systems},
  volume={32},
  year={2019}
}

@inproceedings{AmazonDataset,
author = {He, Ruining and McAuley, Julian},
title = {Ups and Downs: Modeling the Visual Evolution of Fashion Trends with One-Class Collaborative Filtering},
year = {2016},
isbn = {9781450341431},
publisher = {International World Wide Web Conferences Steering Committee},
address = {Republic and Canton of Geneva, CHE},
url = {https://doi.org/10.1145/2872427.2883037},
doi = {10.1145/2872427.2883037},
booktitle = {Proceedings of the 25th International Conference on World Wide Web},
pages = {507–517},
numpages = {11},
location = {Montr\'{e}al, Qu\'{e}bec, Canada},
series = {WWW '16}
}

@inproceedings{unisrec,
  title={Towards universal sequence representation learning for recommender systems},
  author={Hou, Yupeng and Mu, Shanlei and Zhao, Wayne Xin and Li, Yaliang and Ding, Bolin and Wen, Ji-Rong},
  booktitle={Proceedings of the 28th ACM SIGKDD conference on knowledge discovery and data mining},
  pages={585--593},
  year={2022}
}

@inproceedings{peebles2023scalable,
  title={Scalable diffusion models with transformers},
  author={Peebles, William and Xie, Saining},
  booktitle={Proceedings of the IEEE/CVF international conference on computer vision},
  pages={4195--4205},
  year={2023}
}

@inproceedings{bao2023all,
  title={All are worth words: A vit backbone for diffusion models},
  author={Bao, Fan and Nie, Shen and Xue, Kaiwen and Cao, Yue and Li, Chongxuan and Su, Hang and Zhu, Jun},
  booktitle={Proceedings of the IEEE/CVF conference on computer vision and pattern recognition},
  pages={22669--22679},
  year={2023}
}

@inproceedings{bengio2009curriculum,
  title={Curriculum learning},
  author={Bengio, Yoshua and Louradour, J{\'e}r{\^o}me and Collobert, Ronan and Weston, Jason},
  booktitle={Proceedings of the 26th annual international conference on machine learning},
  pages={41--48},
  year={2009}
}

@inproceedings{mu2025trust,
  title={Trust-GRS: A Trustworthy Training Framework for Graph Neural Network Based Recommender Systems Against Shilling Attacks},
  author={Mu, Lingyu and Liu, Zhengxiao and Zhu, Zhitong and Lin, Zheng},
  booktitle={Proceedings of the AAAI Conference on Artificial Intelligence},
  volume={39},
  number={12},
  pages={12408--12416},
  year={2025}
}

@inproceedings{deng2025heterrec,
  title={Heterrec: Heterogeneous information transformer for scalable sequential recommendation},
  author={Deng, Hao and Xing, Haibo and Matsuyama, Kanefumi and Huang, Yulei and Hu, Jinxin and Wen, Hong and Xu, Jia and Chen, Zulong and Zhang, Yu and Zeng, Xiaoyi and others},
  booktitle={Proceedings of the 48th International ACM SIGIR Conference on Research and Development in Information Retrieval},
  pages={3020--3024},
  year={2025}
}

@article{ruc,
  title={LLaDA-Rec: Discrete Diffusion for Parallel Semantic ID Generation in Generative Recommendation},
  author={Shi, Teng and Shen, Chenglei and Yu, Weijie and Nie, Shen and Li, Chongxuan and Zhang, Xiao and He, Ming and Han, Yan and Xu, Jun},
  journal={arXiv preprint arXiv:2511.06254},
  year={2025}
}

@article{baidu,
  title={DiffuGR: Generative Document Retrieval with Diffusion Language Models},
  author={Zhao, Xinpeng and Ren, Zhaochun and Zhao, Yukun and Li, Zhenyang and Zhang, Mengqi and Feng, Jun and Chen, Ran and Zhou, Ying and Chen, Zhumin and Wang, Shuaiqiang and others},
  journal={arXiv preprint arXiv:2511.08150},
  year={2025}
}

@inproceedings{hua2023index,
  title={How to index item ids for recommendation foundation models},
  author={Hua, Wenyue and Xu, Shuyuan and Ge, Yingqiang and Zhang, Yongfeng},
  booktitle={Proceedings of the Annual International ACM SIGIR Conference on Research and Development in Information Retrieval in the Asia Pacific Region},
  pages={195--204},
  year={2023}
}

@inproceedings{lee2022autoregressive,
  title={Autoregressive image generation using residual quantization},
  author={Lee, Doyup and Kim, Chiheon and Kim, Saehoon and Cho, Minsu and Han, Wook-Shin},
  booktitle={Proceedings of the IEEE/CVF conference on computer vision and pattern recognition},
  pages={11523--11532},
  year={2022}
}

@article{hou2024bridging,
  title={Bridging language and items for retrieval and recommendation},
  author={Hou, Yupeng and Li, Jiacheng and He, Zhankui and Yan, An and Chen, Xiusi and McAuley, Julian},
  journal={arXiv preprint arXiv:2403.03952},
  year={2024}
}

@article{lin2025order,
  title={Order-agnostic Identifier for Large Language Model-based Generative Recommendation},
  author={Lin, Xinyu and Shi, Haihan and Wang, Wenjie and Feng, Fuli and Wang, Qifan and Ng, See-Kiong and Chua, Tat-Seng},
  journal={arXiv preprint arXiv:2502.10833},
  year={2025}
}

@article{yang2023diffusion,
  title={Diffusion models: A comprehensive survey of methods and applications},
  author={Yang, Ling and Zhang, Zhilong and Song, Yang and Hong, Shenda and Xu, Runsheng and Zhao, Yue and Zhang, Wentao and Cui, Bin and Yang, Ming-Hsuan},
  journal={ACM computing surveys},
  volume={56},
  number={4},
  pages={1--39},
  year={2023},
  publisher={ACM New York, NY, USA}
}

@String{Computing = "Computing" }

@String{Computer = "{IEEE} Computer" }

@ArtifactSoftware{R,
    title = {R: A Language and Environment for Statistical Computing},
    author = {{R Core Team}},
    organization = {R Foundation for Statistical Computing},
    address = {Vienna, Austria},
    year = {2019},
    url = {https://www.R-project.org/},
}

\end{document}